\documentclass[aps,prl,showpacs,groupedaddress,twocolumn,preprintnumbers,amsmath,amssymb]{revtex4}

\usepackage{graphicx}
\usepackage{dcolumn}
\usepackage{bm}
\usepackage{braket}

\begin{document}

\title{Measuring Qutrit-Qutrit Entanglement of Orbital Angular Momentum States\\of an Atomic Ensemble and a Photon}

\author{R.~Inoue$^{1}$}
\email[]{inoue.r.aa@m.titech.ac.jp}
\author{T.~Yonehara$^{1}$}
\author{Y.~Miyamoto$^{2}$}
\author{M.~Koashi$^{3}$}
\author{M.~Kozuma$^{1,4}$}
\affiliation{%
$^{1}$Department of Physics, Tokyo Institute of Technology, 2-12-1 O-okayama, Meguro-ku, Tokyo 152-8550, Japan}
\affiliation{%
$^{2}$Department of Information and Communication Engineering, The University of Electro-Communications,\\
1-5-1 Chofugaoka, Chofu, Tokyo 182-8585, Japan}
\affiliation{%
$^{3}$Division of Materials Physics, Graduate School of Engineering Science, Osaka University,\\
1-3 Machikaneyama, Toyonaka, Osaka 560-8531, Japan}
\affiliation{%
$^{4}$CREST, Japan Science and Technology Agency, 1-9-9 Yaesu, Chuo-ku, Tokyo 103-0028, Japan}%

\date{\today}
             
\begin{abstract}
Three-dimensional entanglement of orbital angular momentum states of an atomic qutrit and a single photon qutrit has been observed. 
Their full state was reconstructed using quantum state tomography.  The fidelity to the maximally entangled state of Schmidt rank 3 exceeds the threshold $2/3$.
This result confirms that the density matrix cannot be decomposed into ensemble of pure states of Schmidt rank $1$ or $2$. 
That is, the Schmidt number of the density matrix must be equal to or greater than $3$.
\end{abstract}

\pacs{03.65.Wj, 03.67.Mn, 32.80.-t, 42.50.Dv}
\maketitle

An essential requirement toward practical quantum information systems is the capability to generate entangled states between distant sites over quantum networks \cite{kimble_quantum_2008}.  
Inspired by a scheme to create long-lived entanglement in scalable quantum networks proposed by Duan \textit{et al.}~\cite{duan_long-distance_2001}, various experimental results using atomic ensembles have been reported (such as \cite{choi_mapping_2008}).
Recently, d-dimensional quantum systems, or qudits, have been studied, and it has been pointed out that qudits are better adapted for certain purposes. 
As an example, they enable more efficient use of communication channels in quantum cryptography \cite{walborn_2006}.
Following the pioneering experiment on parametric down-conversion \cite{mair_entanglement_2001}, various protocols have been demonstrated using orbital angular momentum (OAM) states of photons (such as \cite{langford_measuring_2004}).  
Photons are a promising carrier of quantum information. 
However, they are difficult to store for appreciable periods of time.  
The realization of massive qudits and the ability to characterize their entanglement are therefore critical for applications.

Another recent landmark is the demonstration of the use of OAM to generate arbitrary superposition of atomic rotational states with the coherent transfer the OAM of light to atoms in Bose-Einstein condensate \cite{andersen_quantized_2006,helmerson_generating_2007}.  
The experiment illustrates the potential of OAM as a tool to generate and to control the atomic qudits.

Previous work \cite{inoue_entanglement_2006} demonstrated entanglement associated with OAM in an ensemble of atoms and a photon. 
The atomic OAM state is linked to the spatial degree of freedom of collective atomic excitations, and, in the case of photons, the OAM state corresponds to Laguerre-Gaussian (LG) modes. 
This result suggests that atomic ensembles can be used as nodes of qudit-based quantum networks. 
However, previous observations were limited to two-qubit entanglement.
In this letter, we report higher-dimensionality of the entanglement of OAM states of an atomic ensemble and a photon, as confirmed by estimating the Schmidt number \cite{terhal_schmidt_2000} of the reconstructed two-qutrit (i.e., qudits with $d=3$) density matrix.
For pure states, the dimension of the range of the marginal state is called the Schmidt rank, which describes how many local levels are involved in the entanglement.  
Extending this notion, the Schmidt number of a bipartite mixed state is defined to be the minimum Schmidt rank that must appear in any decomposition of the state as a mixture of pure states.  
For example, any decomposition of a mixed state with Schmidt number $3$
must include a pure state of Schmidt rank $3$ or greater.

\begin{figure}[b]
\includegraphics[width=86mm]{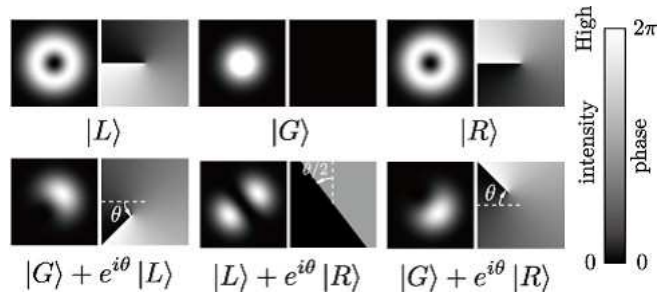}
\caption{\label{fig:lg} The intensity (left panel of each pair) and phase (right panel) distribution of several LG modes and their superpositions. 
}
\end{figure}%
The LG modes constitute a complete basis for describing the paraxial propagation of light \cite{allen_orbital_1992,calvo_quantum_2006}.
The intensity and phase distributions of several LG modes and 
superpositions of them are shown in Fig.~\ref{fig:lg}. 
An LG mode is characterized by its two indices $p$ and $m$, and by the Gaussian beam waist $w_{0}$.
The integers $p$ and $m$ are the radial and azimuthal mode index, respectively, and the phase variation for a closed path around the optical axis is $2m\pi$.
A single photon in the $LG_{p,m}$ mode carries a discrete OAM of $m\hbar$ along its propagation direction.
Here we define $\ket{L}$, $\ket{G}$, and $\ket{R}$ to be the single photon states with OAM of $-\hbar$, $0$, and $+\hbar$, respectively.

\begin{figure}[b]
\includegraphics[width=86mm]{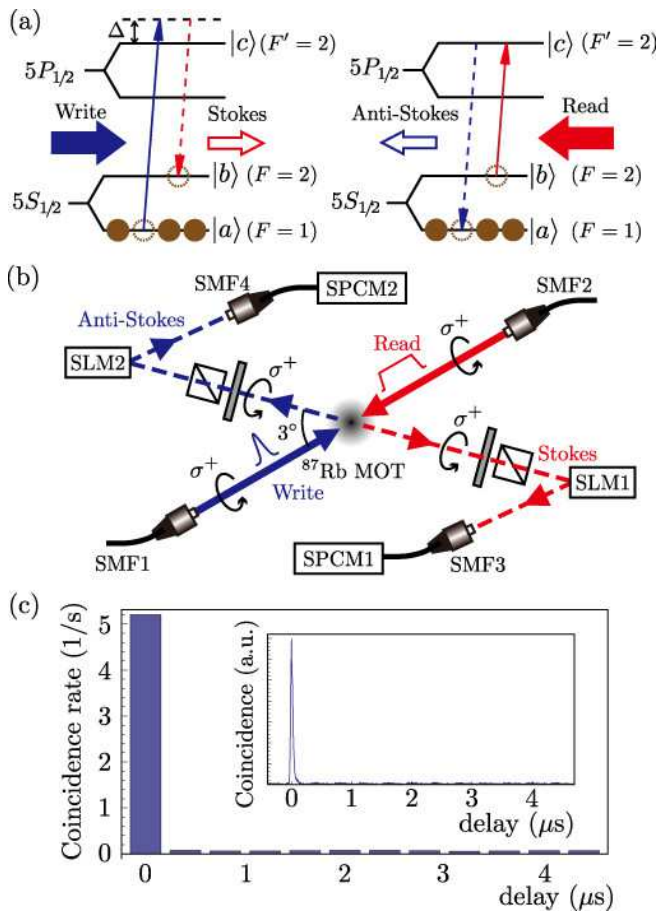}
\caption{\label{fig:ExperimentalSetup} (Color online) (a) Energy levels of ${}^{87}\text{Rb}$, and the associated laser frequencies. (b) Schematic of experimental setup: SLM, spatial light modulator; SMF, single-mode fiber; SPCM, single photon counting module.   Circularly polarized write and read pulses illuminate the ${}^{87}\text{Rb}$ MOT (Magnet-Optical Trap), and circularly polarized Stokes and anti-Stokes photons are selectively directed onto the SPCMs passing through quarter-wave plates and polarizers.  The pair of SLM and SMF serves as a spatial mode filter. (c) Coincidence rate and time-resolved coincidence (inset) between the Stokes and anti-Stokes photons of the $LG_{0,0}$ mode as a function of the time delay.
}
\end{figure}%
Consider an ensemble of atoms having a three-level structure $\ket{a}$, $\ket{b}$, and $\ket{c}$, as shown in Fig.~\ref{fig:ExperimentalSetup}(a). Initially all of the atoms are prepared in level $\ket{a}$.  A classical write pulse tuned to the $\ket{a}\rightarrow\ket{c}$ transition with proper detuning $\Delta$ is incident on the atomic ensemble. In this process, the $\ket{c}\rightarrow\ket{b}$ transition is stimulated and a Stokes photon is generated.  
The write pulse is weak and its interaction time is short. 
Therefore the excitation probability of a Stokes photon into a specified mode is much less than unity per pulse.  
The collective atomic excitation retains the spatial distribution of the relative phase between the write pulse and the Stokes photon.
According to angular momentum conservation, the state of the Stokes photon and collective atomic excitation will be entangled.
We use $\ket{l}$, $\ket{g}$, and $\ket{r}$ to denote the states of the collective atomic excitation with OAM of $-\hbar$, $0$, and $+\hbar$, respectively. 
In the present work, our measurement is sensitive to only the three-dimensional photonic and atomic OAM states.

When the write pulse carries zero OAM, The resultant atoms-photon state will be
\begin{align}\label{eq:state}
\Ket{\phi}=C_{L}\Ket{L}\Ket{r}+C_{G}\Ket{G}\Ket{g}+C_{R}\Ket{R}\Ket{l},\nonumber
\end{align}
where $C_{L}$, $C_{G}$, and $C_{R}$ are the relative complex amplitudes.  The amplitudes depend on the spatial shape of the region where the write pulse interacts with the atoms.



The atoms-photon entanglement can be tested by mapping the state of the atoms to that of an anti-Stokes photon by illuminating the atomic ensemble with a laser pulse (read pulse) resonant with the $\ket{b}\rightarrow\ket{c}$ transition.
The efficiency of this transfer can be nearly unity because it corresponds to the retrieval process of the atomic quantum memory based on electromagnetically induced transparency \cite{fleischhauer_quantum_2002}.


A schematic of the experimental setup is shown in Fig.~\ref{fig:ExperimentalSetup}(b).  
An optically thick (optical depth of about 5) cold atomic cloud is created using a magneto-optical trap (MOT) for $^{87}\mathrm{Rb}$.  
Levels $\ket{a}$ and $\ket{b}$ correspond to $5S_{1/2}F=1$ and $2$, respectively, and level $\ket{c} $ corresponds to $5P_{1/2}F^\prime=2$. 
One cycle in the experiment comprises a $6$-ms loading period and a $4$-ms measurement period.
During the loading period, a gas of cold $^{87}\mathrm{Rb}$ atoms is cooled and trapped, and they are optically pumped to the $5S_{1/2}F=1$ level using a $50\-\mathrm{\mu s}$ depumping pulse tuned to the $5S_{1/2}F=2\rightarrow 5P_{3/2}F^\prime=2$ transition.
After the loading period, the magnetic field and the radiation responsible for the cooling, trapping, and depumping are shut off.  
The vacuum cell is magnetically shielded using a single-layer permalloy.  
The coil jig is non-metallic, and thus eddy currents, which prolong the decay of the magnetic field \cite{inoue_entanglement_2006}, are suppressed. The residual magnetic field is about $1$ mG during the measurement period.
During that period, read and write pulses illuminate the atomic ensemble with a $400$-ns repetition cycle.  
The $15$-ns Gaussian write pulse is tuned to the $\ket{a}\rightarrow\ket{c}$ transition with $10$-MHz detuning and comprises $4\times 10^4$ photons.
After a $40$-ns delay, a $200$-ns rectangular read pulse of $300\-\mathrm{\mu W}$ intensity illuminates the MOT.
As shown in Fig.~\ref{fig:ExperimentalSetup}(b), the read and write pulses, whose spatial modes are cleaned up by passage through single-mode fibers SMF1 and SMF2, are counter-propagated along the same axis.  
The output beam from the fiber is focused into the MOT with a Gaussian beam waist of $400\ \mathrm{\mu m}$.
Similarly, the Stokes and anti-Stokes photons share a single spatial mode when their conversions are appropriately chosen using spatial light modulators SLM1 and SLM2 (Hamamatsu model X8267).
Their Gaussian beam waist at the center of the MOT is adjusted to $275\,\mathrm{\mu m}$. 
The angle between the axis of the Stokes or anti-Stokes photons and that of SMF1 or SMF2 is $\sim 3^\circ$ in order to spatially separate the weak Stokes (anti-Stokes) photons from the strong write (read) pulses.
The incident write and read pulses are circularly polarized, as are the Stokes and anti-Stokes photons.

The Stokes and anti-Stokes photons coupled into SMF3 and SMF4 are directed onto the single-photon-counting modules SPCM1 and SPCM2 (Perkin-Elmer model SPCM-AQR-14). 
From the Stokes photon counting, the excitation probability is estimated to be $5\times10^{-4}$.
Their outputs are then fed into the start and stop inputs of the time interval analyzer.
The experimental results for the coincidence rate between the Stokes and anti-Stokes photons in an $LG_{0,0}$ mode are displayed versus time delay in Fig.~\ref{fig:ExperimentalSetup}(c), from which the normalized cross-intensity correlation is estimated to be $g^{(2)}_\text{s,as}=74.6\pm7.4$, confirming that the excitation probability of a Stokes photon into a specified mode is much less than unity for each pulse.
The coincidence count rate was $5.2\ \mathrm{s}^{-1}$.

\begin{figure}[b]
\includegraphics[width=86mm]{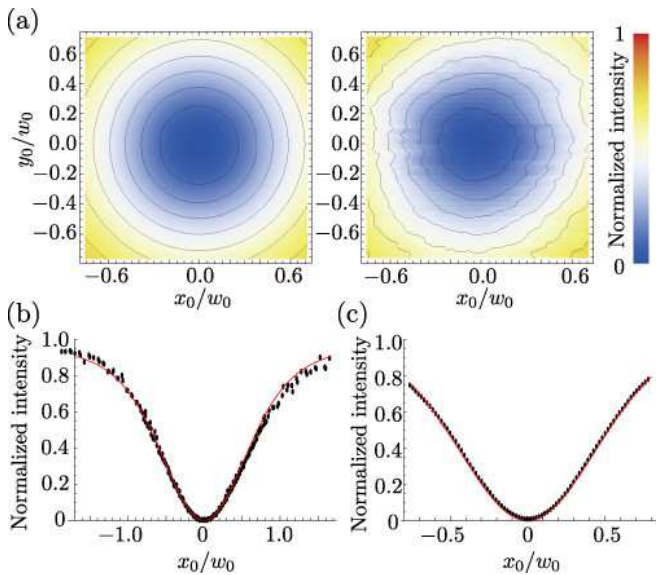}
\caption{\label{fig:slm} (Color online) 
Gaussian components of applied phase modulations $T(x,y)$ to an incoming Gaussian beam of waist $w_{0}=2.2\ \mathrm{mm}$. 
(a) $T(x,y)=e^{i\text{arg}(x-x_{0}+i(y-y_{0}))}$.  The left panel plots the simulation while the right panel shows the experimental result. 
(b) $T(x,y)=e^{i\text{arg}(x-x_{0}+iy)}$. 
(c) $T(x,y)=e^{i\frac{\pi}{2}\text{sgn}(x-x_{0})}$.
In panels (b) and (c), the dots are experimental results and the solid curves are obtained from the numerical simulation.
}
\end{figure}%

To determine the full state of the atoms and Stokes photon, two-qutrit state tomography \cite{james_measurement_2001,thew_qudit_2002} was performed, where the density matrix was reconstructed from the set of $81$ measurements represented by the operators $\hat{\mu}_{i}\otimes\hat{\mu}_{j}$ (with $i,j=0,1,\cdots,8$) and where $\hat{\mu}_{k}\equiv\Ket{k}\!\!\Bra{k}$.  
The ket $\Ket{k}$ for the Stokes photon was chosen from among
$\{\Ket{L},\ \Ket{G},\ \Ket{R},\ (\Ket{G}+\Ket{L})/\sqrt{2},\ (\Ket{G}+\Ket{R})/\sqrt{2},\ (\Ket{G}+i\Ket{L})/\sqrt{2},\ (\Ket{G}-i\Ket{R})/\sqrt{2},\ (\Ket{L}+\Ket{R})/\sqrt{2},\ (\Ket{L}+i\Ket{R})/\sqrt{2}\}$, and for the collective atomic excitation from among $\{\Ket{l},\ \Ket{g},\ \Ket{r},\ (\Ket{g}+\Ket{l})/\sqrt{2},\ (\Ket{g}+\Ket{r})/\sqrt{2},\ (\Ket{g}-i\Ket{l})/\sqrt{2},\ (\Ket{g}+i\Ket{r})/\sqrt{2},\ (\Ket{l}+\Ket{r})/\sqrt{2},\ (\Ket{l}-i\Ket{r})/\sqrt{2}\}$.
These measurements are implemented using SLMs and SMFs; the SLMs produce spatial phase modulation and the SMFs filter the $LG_{0,0}$ mode \cite{yao_observation_2006}.
As reported in \cite{vaziri_superpositions_2002} in detail, arbitrary superpositions of an $LG_{0,0}$ and an $LG_{0,\pm1}$ mode can be converted into an $LG_{0,0}$ mode by application of the phase modulation $T$ corresponding to the relative phase difference between the superposition mode and the $LG_{0,0}$ mode.
As is clear from Fig.~\ref{fig:lg}, such conversions can be achieved by moving the singularity in the phase modulation to a particular location.  
Our reflective SLMs have an active region of $768\,\mathrm{px}\times768\,\mathrm{px}$.  
Even if the phase modulation is discrete, the fractional intensity diffracted into higher orders can be decreased by adding the blazed phase grating structure.  
The spatial period of the grating is $4\ \mathrm{px}\sim100\ \mathrm{\mu m}$ with a diffraction efficiency of $25\ \%$. 
In order to check the SLMs, the Gaussian components of a beam diffracted with a spatial phase modulation of $T(x,y)=e^{i\text{arg}(x-x_{0}+i(y-y_{0}))}$ were measured. 
Here $\text{arg}(z)$ is the argument of the complex number $z$.
The results are plotted in Fig.~\ref{fig:slm}(a) and (b), and are in good agreement with numerical calculations of the superposition modes.
The location of the singularity that converts the superposition mode into a Gaussian can therefore be determined.  
At position $(x_{0},y_{0})=(0,0)$, where an incoming Gaussian mode is converted into an $LG_{0,\pm1}$ mode, the normalized intensity was measured to be $3\times10^{-3}$, indicating a high extinction ratio.
Similarly, the superposition mode $(LG_{0,-1}+e^{i\theta}LG_{0,+1})/\sqrt{2}$ can be converted into an $LG_{0,0}$ mode by applying a discontinuous phase modulation.   
The Gaussian components of the beam diffracted by an SLM with a spatial phase modulation of $T(x,y)=e^{i\frac{\pi}{2}\text{sgn}(x-x_{0})}$ was also measured, where $\text{sgn}(x)$ is the sign of $x$.  
The experimental results are shown in Fig.~\ref{fig:slm}(c), and are again in good agreement with numerical calculations.

\begin{figure}[tbp]
\includegraphics[width=86mm]{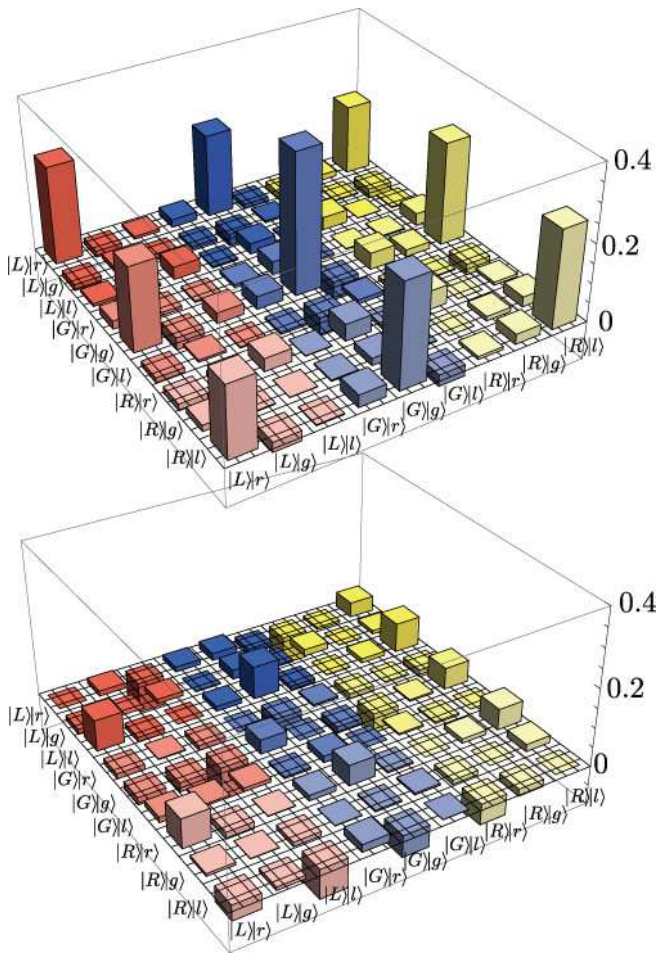}
\caption{\label{fig:DensityMatrix}
(Color online) Graphical representation of the density matrix $\rho_\text{exp}$ of a state as estimated by quantum state tomography from the experimentally obtained coincidences.  The upper plot is the real part, and the lower plot is the imaginary part.
}
\end{figure}

Fig.~\ref{fig:DensityMatrix} shows the graphical representation of density matrix $\rho_\text{exp}$ reconstructed from the $81$ coincidences.  
The typical coincidence rate was roughly $5\ \mathrm{s}^{-1}$, and the data acquisition time of each measurement was $100$ s. 
From the density matrix, the fidelity to a maximally entangled state $F_\text{exp}\equiv\Braket{\text{MES}|\hat{\rho}_\text{exp}|\text{MES}}=0.74\pm0.02$ was obtained.
Here $\Ket{\mathrm{MES}}$ was chosen from the set of maximally entangled states $(e^{i\alpha\pi}\ket{L}\ket{r}+\ket{G}\ket{g}+e^{i\beta\pi}\ket{R}\ket{l})/\sqrt{3}$ so as to maximize the fidelity, where the values of $\alpha$ and $\beta$ were $0.019\pi$ and $-0.058\pi$, respectively.
The error in $F_\text{exp}$ is calculated by using Monte-Carlo method from the statistical uncertainties in the coincidences. 
An optimal witness operator of Schmidt number 3 in $C^3\otimes C^3$ is given by $\hat{W}_{3}=1-3\Ket{\text{MES}}\!\!\Bra{\text{MES}}/2$ \cite{sanpera_schmidt-number_2001}, resulting in $\text{Tr}(\hat{W}_{3}\hat{\rho})<0\Leftrightarrow \Braket{\text{MES}|\hat{\rho}|\text{MES}}>2/3$.
The experimental result $F_\text{exp}>2/3$ therefore confirms that the Schmidt number of the mixed state of the atomic ensemble and the photon is greater than or equal to $3$.

The OAM measurements were achieved using mode conversion by SLMs and mode filtering by SMFs in this experiment.
However, as frequently occurs in experiments utilizing a spatial phase modulation, the measurement bases cannot be realized completely accurately, resulting in unwanted radial and azimuthal components comprising up to $20{\%}$.
While this systematic effect increases the error in $F_\text{exp}$, the resultant fidelity is nevertheless $0.74^{+0.06}_{-0.07}$, 
which is larger than $2/3$ even at the lowest error bound.

The major diagonal elements of the reconstructed density matrix are  
$\bra{L}\!\bra{r}\!\rho_\text{exp}\!\ket{L}\!\ket{r}=0.25,\ \bra{G}\!\bra{g}\!\rho_\text{exp}\!\ket{G}\!\ket{g}=0.37,\ \text{and }\bra{R}\!\bra{l}\!\rho_\text{exp}\!\ket{R}\!\ket{l}=0.26$. 
The summation of remaining elements $1-0.25-0.37-0.26=0.12$ suggests that components of non-zero total OAM, which may originate mainly from stray light, are one of the dominant factors limiting the fidelity.
The normalized cross intensity correlation $g^{(2)}_\text{s,as}=74.6\pm7.4$ is significantly smaller than the value $2000$ expected from the measured excitation probability of $5\times10^{-4}$.  
Therefore, stray light appears to have strongly affected the photon statistics and decreased the fidelity.
The imbalance of the diagonal elements may also affect the fidelity.  
However, even supposing that the Stokes photons are locally filtered so as to balance the relative amplitude, the fidelity is still expected to be $0.74$.
This result confirms that the imbalance is not the dominant factor decreasing the fidelity in our case.
Note that the diagonal elements can be balanced by changing parameters such as the beam waist of the write pulse since the relative amplitude is dependent on the spatial shape of the effective interaction volume \cite{osorio_spatial_2008}. 
The other factor limiting the fidelity is the decoherence caused by the environmental noise, such as the Larmor precession of the ground-state Zeeman sublevels and the ballistic expansion of the atomic ensemble.

In conclusion, higher-dimensionality of the entanglement of OAM states has been observed for an atomic ensemble and a photon by estimating the Schmidt number of the reconstructed density matrix. 
The experiment described here enables one to communicate quantum information encoded in the spatial degrees of freedom of a photon and an atomic ensemble \cite{molina-terriza_twisted_2007}.

We gratefully acknowledge M.~Ueda, K.~Usami, and N.~Kanai for valuable comments and stimulating discussions.
R.I. was supported by a Grant-in-Aid from JSPS.  
This work was supported by the Global Center of Excellence Program by MEXT, Japan through the Nanoscience and Quantum Physics Project of the Tokyo Institute of Technology, and by MEXT through a Grant-in-Aid for Scientific Research (B).


\begin{thebibliography}{20}
\bibitem{kimble_quantum_2008}H.~J. Kimble, Nature {\bf 453},  1023  (2008).
\bibitem{duan_long-distance_2001}L. Duan, M.~D. Lukin, J.~I. Cirac, and P. Zoller, Nature {\bf 414}, 413 (2001).
\bibitem{choi_mapping_2008}K.~S. Choi, H. Deng, J. Laurat, and H.~J. Kimble, Nature {\bf 452}, 67 (2008).
\bibitem{walborn_2006}S.~P.~Walborn, D.~S.~Lemelle, M.~P.~Almeida, and P.~H.~Souto~Ribeiro, Phys.~Rev.~Lett. {\bf 96}, 090501 (2006).
\bibitem{mair_entanglement_2001}A. Mair, A. Vaziri, G. Weihs, and A. Zeilinger, Nature {\bf 412}, 313 (2001).
\bibitem{langford_measuring_2004}N.~K. Langford {\it et~al.}, Phys.~Rev.~Lett. {\bf 93}, 053601 (2004).
\bibitem{andersen_quantized_2006}M.~F.~Andersen {\it et~al.}, Phys. Rev. Lett. {\bf 97}, 170406 (2006).
\bibitem{helmerson_generating_2007}K. Helmerson {\it et~al.}, Nucl. Phys. A, {\bf 790}, 705c-712c (2007).
\bibitem{inoue_entanglement_2006}R. Inoue {\it et~al.}, Phys.~Rev.~A {\bf 74}, 053809 (2006).
\bibitem{terhal_schmidt_2000}B.~M. Terhal and P. Horodecki, Phys.~Rev.~A {\bf 61}, 040301 (2000).
\bibitem{allen_orbital_1992}L. Allen, M.~W. Beijersbergen, R.~J.~C. Spreeuw, and J.~P. Woerdman, Phys.~Rev.~A {\bf 45}, 8185  (1992).
\bibitem{calvo_quantum_2006}G.~F. Calvo, A. Picon, and E. Bagan, Phys.~Rev.~A {\bf 73}, 013805 (2006).
\bibitem{fleischhauer_quantum_2002}M. Fleischhauer and M.~D. Lukin, Phys.~Rev.~A {\bf 65}, 022314 (2002).
\bibitem{james_measurement_2001}D.~F.~V. James, P.~G. Kwiat, W.~J. Munro, and A.~G. White, Phys.~Rev.~A {\bf 64}, 052312 (2001).
\bibitem{thew_qudit_2002}R.~T. Thew, K. Nemoto, A.~G. White, and W.~J. Munro, Phys.~Rev.~A {\bf 66}, 012303 (2002).
\bibitem{yao_observation_2006}E.~Yao {\it et~al.}, Optics Express {\bf 14}, 13089 (2006).
\bibitem{vaziri_superpositions_2002}A.~Vaziri, G.~Weihs, and A.~Zeilinger, Journal of Optics B: Quantum and Semiclassical Optics {\bf 4}, S47 (2002).
\bibitem{sanpera_schmidt-number_2001}A.~Sanpera, D.~Bru{\ss}, and M.~Lewenstein, Phys.~Rev.~A {\bf 63}, 050301(R) (2001).
\bibitem{osorio_spatial_2008}C.~I. Osorio, S. Barreiro, M.~W. Mitchell, and J.~P. Torres, Phys.~Rev.~A {\bf 78}, 052301 (2008).
\bibitem{molina-terriza_twisted_2007}G.~{Molina-Terriza}, J.~P. Torres, and L. Torner, Nature Physics {\bf 3}, 305 (2007).
\end{thebibliography}
\end{document}